\documentclass{ws-mpla}

\newcommand{\minerva}{\mbox{MINERvA}}

\begin{document}

\markboth{ }
{DEMONSTRATION OF COMMUNICATION USING NEUTRINOS}

\catchline{}{}{}{}{}

\title{DEMONSTRATION OF COMMUNICATION USING NEUTRINOS}

\author{D.D.~Stancil$^1$, P.~Adamson$^2$, M.~Alania$^3$, L.~Aliaga$^4$, M.~Andrews$^2$, C.~Araujo Del Castillo$^4$, L.~Bagby$^2$, J.L.~Bazo Alba$^4$, A.~Bodek$^5$, D.~Boehnlein$^2$
% ...
}

\address{ {$^1$Department of Electrical and Computer Engineering, North Carolina State University, Raleigh, NC  27695,\\$^ 2$Fermi National Accelerator Laboratory, Batavia, IL  60510,  \\$^3$Universidad Nacional de Ingenieria, Av. Tupac Amaru 210, Lima, Peru, \\$^4$Secci\'{o}n F\'{i}sica, Departamento de Ciencias, Pontificia Universidad 
Cat\'{o}lica del Peru, Apartado 1761, Lima, Peru, \\$^5$Department of Physics and Astronomy, University of Rochester, Rochester NY 14627. \\
% ...
} 
\\
{ddstancil@ncsu.edu} }



\maketitle

\begin{abstract}
Beams of neutrinos have  been proposed as a vehicle for 
communications under unusual circumstances, 
such as direct point-to-point global communication, communication 
with submarines, secure communications and interstellar communication. 
We report on the performance of a low-rate communications link established 
using the NuMI beam line and the \minerva\ detector at Fermilab. 
The link achieved a decoded data rate of 0.1 bits/sec with a bit error rate 
of 1\% over a distance of 1.035 km, including 240 m of earth. 

\end{abstract}

\ccode{PACS Nos.: 14.60.Lm, 01.20.+x}



The use of fundamental particles that interact very weakly with matter
has been proposed as a vehicle for communication in environments where
electromagnetic waves are damped and do not penetrate easily. The bulk
of these proposals have involved
neutrinos,\cite{Arnold,Saenz,Subotowicz,Learned,Zee,huber} although
recent consideration has also been given to possible use of
hypothetical particles such as the axion\cite{Stancil} and the
hidden-sector photon.\cite{Jaeckel}  Neutrino communication systems
have been proposed for point-to-point global
communications,\cite{Saenz} global communication with
submarines,\cite{Saenz,huber} and for interstellar
communication.\cite{Subotowicz,Learned} Neutrino communication systems
can also be contemplated for use in planetary exploration during
periods in which the communications link is blocked by a planetary
body. Even low bandwidth global point-to-point links could be useful
for secure exchange of encryption codes. While the ability to
penetrate matter is an important advantage of neutrinos, the weak
interactions of neutrinos also imply that very intense beams and
massive detectors would be required to realize this type of
communication.

We report on the performance of a low-rate communications link
established using the NuMI beam line and the \minerva\ detector at
Fermilab. The link achieved a decoded data rate of 0.1 bits/sec with a
bit error rate of 1\% over a distance of 1.035 km that included 240 m
of earth. This demonstration illustrates the feasibility of using
neutrino beams to provide a low-rate communications link, independent
of any existing electromagnetic communications infrastructure.
However, given the limited range, low data rate, and extreme
technologies required to achieve this goal, significant improvements
in neutrino beams and detectors are required for ``practical''
application.

\begin{figure}[htbp]
\centerline{\psfig{file=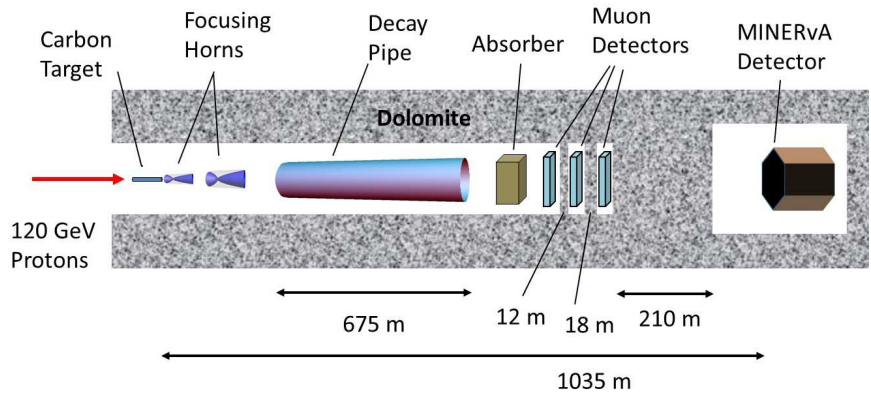,width=4.5in}}
\vspace*{8pt}
\caption{Layout of the NuMI beam line used as the neutrino source, and the \minerva\ detector.  \protect\label{fig1}}
\end{figure}

The Neutrino beam at the Main Injector (NuMI)\cite{numi} at Fermilab
is currently one of the most intense high energy neutrino beams
worldwide, and provides a variable-energy beam for use in particle
physics experiments. A simplified schematic of the beam line is shown
in Fig.~\ref{fig1}.  A series of accelerators currently produces 8.1
$\mu$s pulses of 120 GeV protons every 2.2 s.  The repetition rate is
limited by the time required to accelerate the protons to high energy.
The proton beam strikes a carbon target that produces many pions,
kaons, and other particles. Charged particles are focused using
magnets to produce a beam that is directed toward detectors. Almost
all pions and kaons decay into
neutrinos (mostly muon neutrinos and other associated particles) in a 675~m helium-filled decay pipe, in roughly the
same direction as the momentum of the original meson.  This beam of
assorted elementary particles passes through 240 meters of rock
(mostly shale), and all particles are absorbed, except the neutrinos
that define the NuMI beam. Since proton-carbon interactions at 120 GeV
produce pions much more readily than kaons, the beam is predominantly
composed of muon neutrinos (88\%) with smaller components of muon
antineutrinos (11\%) and electron neutrinos (1\%).  The energy
spectrum peaks at 3.2 GeV, and has a width (FWHM) of about 2.8 GeV.
Although the energies extend to 80 GeV, 91.6\% of the
neutrinos in the beam have energies below 10 GeV.  The detector is slightly more
than 1 km from the target, and, at its peak energy, the beam has a
transverse size of a few meters at the face of the detector.

\begin{figure}[htbp]
\centerline{\psfig{file=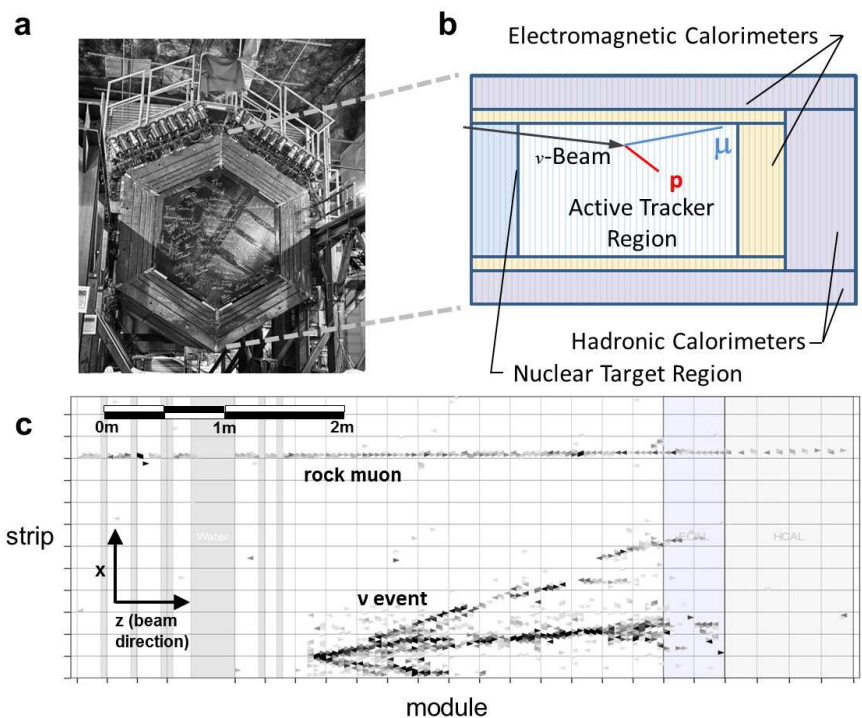,width=4.5in}}
\vspace*{8pt}
\caption{The \minerva\ detector. (a) Photograph. (b) Cross sectional
  diagram of the detector (c)
  Events in the detector from a single beam pulse that 
  contains both kinds of events (neutrino interactions in the target and in the rock before the target) considered in this
  measurement. Each pixel contains position and energy
  information, and the display shows just the central region of the tracker (only the central portion was used).
  \protect\label{fig2}}
\end{figure}

The \minerva\ detector (Fig.~\ref{fig2}) is located in a cavern about
100 m underground.  Cross sections for neutrino interactions are
measured by observing the trajectories of particles emitted when
neutrinos interact with atomic nuclei in the material of the detector
(carbon, lead, iron, water, helium and scintillator).  The basic
element of the detector is a hexagonal plane assembled from parallel triangular
scintillating strips.  The full detector has 200 such planes, and a
total weight of 170 tons.  The detector planes have one of three
orientations rotated from each other by 60 degrees, and are labeled X,
U, or V depending on the orientation.  Analyses in the
\minerva\ experiment focus on neutrino interactions in the central
tracker, comprising a three-ton detector that is fully sensitive to
particles produced in the neutrino collisions.  A charged particle
passing through a scintillator produces scintillation light proportional
to its deposited energy.  When the light passes through wavelength
shifting fibers embedded in the scintillator, fluorescent dopants in
the fibers emit light at a new (green) wavelength, much of which is
then transmitted through the optical fibers.  Optical cables transport
that light to photomultiplier tubes located above
the detector. These signals are used to determine the deposited energy of the particle and the position of the
particle in two dimensions.  In turn, this information
is used to reconstruct particle identities and their trajectories in three dimensions by combining the X, U and V views to
determine the products of the neutrino interaction.  The central
tracker is surrounded by calorimeters that have alternating layers of
metal and scintillator used to contain high energy showering particles
and measure their energy.  The detector was designed to collect more than 16 million
neutrino events over four years of beam time.  

The signal for this measurement is charged-current neutrino interactions that contain muons in the
final state.  The range of these muons is typically
many meters in the plastic scintillator, so that each muon leaves a signal
in at least several tens of scintillator strips, and such events are
distinguished by their long and straight trajectories.   The efficiency for
reconstructing these long muon tracks is better than 95\%, and there is a
very small probability, negligible for the purposes of this demonstration, of finding
such a muon candidate with the beam off. 
  
In this demonstration, most of the signal is from neutrino
interactions in the rock upstream of \minerva\ that produce muons that
enter the upstream face of the detector.  A smaller
component of the signal is from neutrino interactions that produce muons in the
active region of the detector. 
At the reduced intensity of $2.25
\times 10^{13}$ protons per pulse used in this study of neutrino
communication, an average of 0.81 events is registered during each
pulse, through the use of a software filter to identify these two
classes of signal muons that indicate the beam is ``on''.  With a time resolution
of a few nanoseconds for the muons selected for this demonstration of principle, 
it is possible to disentangle
multiple events in the 8.1~$\mu$s beam bursts.  An example of multiple
events resulting from a single beam pulse is shown in Fig.~3.
Figure 4 shows a histogram of the number of muons seen per beam pulse
when the beam is not modulated.  The close correspondence between
data and a Poisson distribution shows the uncorrelated nature of our
events, which is assumed in the statistical analysis of the communications
experiment.

\begin{figure}[htbp]
\centerline{\psfig{file=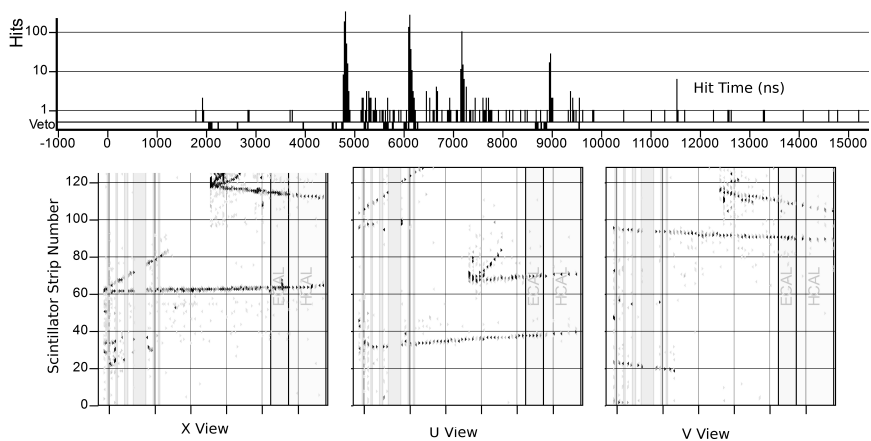,width=4.5in}}
\vspace*{8pt}
\caption{Example of track images from the \minerva\ detector.  The
  upper panel shows 6 separate events occurring at different times
  during an 8.1~$\mu$s neutrino pulse. The time measurements
  separate them easily.  A superposition of tracks from these 6 events
  is shown for the X, U, and V planes as a function of module number
  in the stack in the lower panels.  The horizontal axis is distance
  (about 3.6 cm/module) along the axis of the central tracker, and the
  vertical axis is distance (about 1.8 cm per strip) in directions
  perpendicular to the central axis.  
  \protect\label{fig3}}
\end{figure}

\begin{figure}[htbp]
\centerline{\psfig{file=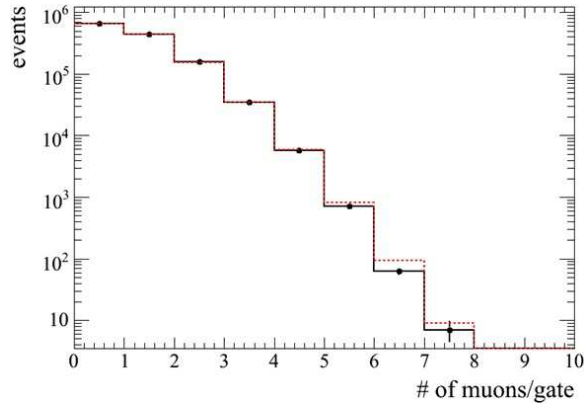,width=3in}}
\vspace*{8pt}
\caption{Muons per beam pulse in a run without beam modulation and higher beam
  intensity than available for this experiment. The black
  histogram is data and the dotted histogram is a Poisson
  distribution for the appropriate mean. The statistical error bars are
  in most cases smaller than the data points.
  \protect\label{fig4}}
\end{figure}

The simplest method for encoding information about the neutrino beam
is on-off keying (OOK). In this scheme, a ``1'' or a ``0'' is
represented by the presence or absence of a beam pulse,
respectively. In the NuMI beam line, OOK was implemented by
controlling proton beam pulses from the Main Injector.  Because the
beam has a specific pulsed time structure, and because there are many independent
detector segments, the chance of seeing an event that is not caused by
the NuMI beam, e.g., from cosmic rays, is extremely small. A ``1'' bit
corresponds to a beam pulse with an observed event or events, and a
``0'' bit is a pulse with no events.

If the occurrence of events is modeled as a random Poisson process,
this channel is mathematically equivalent to the photon counting (or
Poisson) channel in communication theory\cite{Pierce} that models
direct-detection of optical communications. In this model, the
probability that no events are observed during a pulse is
$e^{-\lambda}$, where $\lambda$ is the expected number of events per
pulse when the beam is ``on''. Since the neutrino detection system
produces no errors when a 0 is sent, the probability of a bit error,
or bit-error rate (BER), is $e^{-\lambda}/2$, assuming that 1 and 0
are equally likely.

The structure of the message is shown in Fig.~5a. The 8-character
word ``neutrino'' is expressed in an abbreviated 5-bit code obtained
by omitting the first two (left-most) bits in the standard 7-bit American Standard Code for Information Interchange (ASCII) code. This 40-bit
message is subsequently encoded using the convolutional code with rate
$\frac{1}{2}$ and constraint length 7, that is, in the NASA/ESA
Planetary Standard,\cite{Costello} and expands the message to 92
bits. This encoded message is concatenated with a 64-bit pseudo-noise
(PN) synchronization sequence to form a 156-bit frame. This frame is
repeated for the duration of the communication experiment. The
accelerator throughout this study was operated at 25 pulses spaced by
2.2 s, followed by a 6.267 s interval to form a 61.267 s
``supercycle.''

\begin{figure}[htbp]
\centerline{\psfig{file=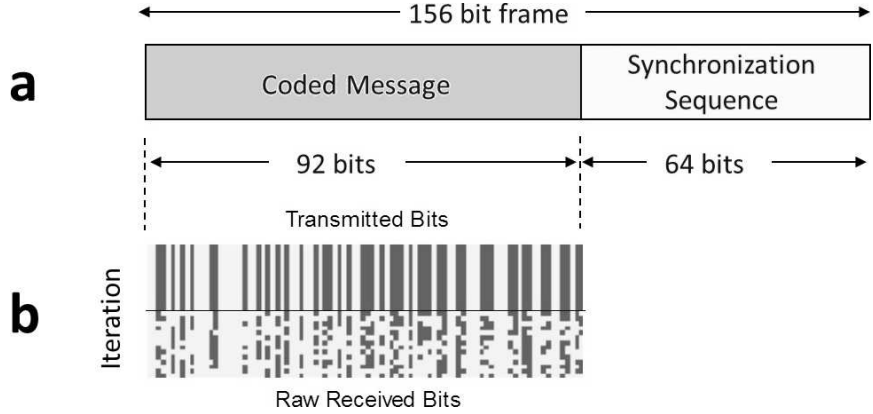,width=4.5in}}
\vspace*{8pt}
\caption{ Format and representations of transmitted data. (a) Frame structure of the transmission. (b) Depictions of the transmitted and received data.  \protect\label{fig5}}
\end{figure}

The data received consists of 3454 records spanning an interval of
just over 142 minutes, along with time stamps for each record. The
detector is read out whenever a timing gate is received, independent
of the presence of any beam. Each record contains the number of events
observed in the detector over the duration of a beam pulse. Decoding
the message requires the locations of the frames
within the data. This is done by 
searching for the 64-bit PN sequence as a ``sync'' word that signals
the end of each frame. If $x_0,\ldots, x_{N-1}$ is the entire sequence
of transmitted bits and $K_0,\ldots, K_{N-1}$ the corresponding
received counts, then $K_i$ should be a Poisson random variable with
parameter $\lambda x_i + \lambda_0$, where $\lambda$ is the expected
number of events from a transmitted pulse, and $\lambda_0 \ll
\lambda$ is the number of background events, which is nearly zero in this
test. If the counts are conditionally independent, given $x_0,\ldots,
x_{N-1}$, the receiver model is then mathematically equivalent to the
direct-detection optical channel.\cite{liu} We searched for the end of
a known sync word, e.g., $S_0,\ldots, S_{L-1}$, by finding the maxima of
the log-likelihood statistic
\begin{equation}
 \Lambda (m) = \sum^{L-1}_{j=0} \left( S_{L-1-j} - \frac{1}{2} \right) K_{m-j}.
 \label{eqn1}
 \end{equation}
However, to ensure symbol synchronization throughout the frame we
search for two sync words spaced exactly by $m_0 = 156$ counts,
so that the test statistic becomes $\Gamma (m) = \Lambda (m) + \Lambda
(m-m_0)$.  At the end of a sync word, this statistic has a mean $\mu =
L\lambda/2$ and variance $\sigma^2 = L\lambda/2$. To detect a frame,
we looked for $m $ such that $\Gamma (m) \geq \mu - 2\sigma$.  If half
the bits are zero and $\lambda_0 \approx 0$, an estimate of $\lambda$
is obtained by noting that 1402 events are observed in 3454 counts, so that 
$\lambda \approx 2 \times 1402/3454 = 0.81$.  As illustrated in Fig.~6, this statistic identifies 15 frames with symbol synchronization,
indicated by open circles. There are clearly other frames present that might be decoded successfully with more advanced processing;
nevertheless, we will not pursue this here.

\begin{figure}[htbp]
\centerline{\psfig{file=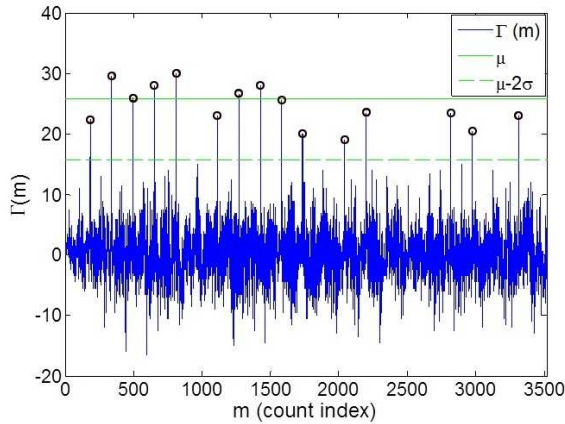,width=3.0in}}
\vspace*{8pt}
\caption{Statistic used for synchronization of frames.  The open circles identify the locations of correctly reconstructed frames.  Some frames were not correctly found because of accelerator aborts or data acquisition deadtimes.
  \protect\label{figZZZ6}}
\end{figure}

The 15 synchronized frames are used to recover the message and to
estimate the properties of the channel. For any nonzero count, the bit
is estimated to be 1, and 0 otherwise. Figure 5b illustrates message
recovery in pictorial form, with each row of pixels in the top half of
the image representing the transmitted data in one of the 15 frames,
where dark pixels represent ones and light pixels are zeros. The
bottom half of Fig.~5b shows the corresponding estimated bits at the
receiver, where most (78\%) are received correctly. Errors can be
reduced by combining multiple frames and by using the convolutional
error-control code. The optimal way to combine frames is to add the
corresponding event counts. In fact, the transmitted message is
recovered perfectly if the results of all the frames are pooled, since
events are observed in each column where a ``1'' is transmitted, and
no events occur in any column where a ``0'' is
transmitted. Consequently, this result demonstrates the ability to
communicate information using neutrinos.

Figure 7 illustrates the relationship between the number of frames
combined and the uncoded bit-error rate (BER) of the channel,
which is the average fraction of incorrect received bits. Also shown
for comparison is the theoretical uncoded BER of the Poisson
prediction of ${\rm\textstyle BER} = e^{-\lambda}/2$ for $\lambda = 0.81$.  The
measured results are in close agreement (within expected fluctuations)
with the Poisson model.  It should be noted that 99\% of the
transmitted bits are decoded correctly using only 5 frames. No errors
were observed for any combination of 9 or more frames.

Figure 7 also shows the BER after error-control decoding with the
Viterbi algorithm.\cite{Costello} These values are in very good
agreement with the simulated Poisson channel for $\lambda = 0.81$
for the same type of decoding. In this case, $\approx$99\% of the
transmitted bits are received correctly after only two frames, and no
decoding errors are observed when any three frames are combined.

\begin{figure}[htbp]
\centerline{\psfig{file=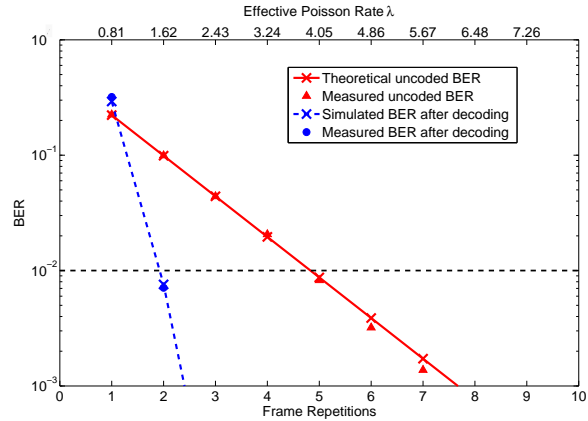,width=3.5in}}
\vspace*{8pt}
\caption{A comparison of predicted and measured bit-error rates, with
  and without error-control coding.}
\label{fig7}
\end{figure}

The theoretical capacity of the photon counting channel with OOK
modulation and background rate $\lambda_0 = 0$ is given
by\cite{Bar-David}
\begin{equation}
C/B = \log_2 \left[ 1 + (1-e^{-\lambda}) \cdot \exp \left( - \frac{\lambda}{e^\lambda - 1}\right)\right]  \;\; ({\rm bits/pulse}),
\label{eqn2}
\end{equation}
where $B$ is the pulse or baud rate (pulses/sec), $C$ is the channel
capacity (bits/s), and $\lambda$ is the average number of events per
pulse defined above. In our experiment $\lambda \approx 0.81$,
which yields a theoretical capacity of $C/B \approx 0.37$ bits/pulse. For
comparison, the message was decoded with a 1\% error rate after two
iterations, or for the transmission of $ 2 \times 92 = 184$ pulses. Since
there were only 40 bits of information transmitted, the experimental
data rate can be estimated as 0.22 bits/pulse, or more than half the
theoretical capacity of 0.37 bits/pulse. Since the pulse rate was
about 1/2.2 bits per second, the experimental data rate at an error
probability of 1\% can be estimated as $0.22\times (1/2.2) \approx
0.1$ bits per second.  We stress that this is only a rough estimate,
since there are only 105 combinations of 2 frames selected from 15,
and these combinations do not represent independent measurements
because of reuse of the frames in forming the combinations.

In general, long-distance communication using neutrinos will favor
detectors optimized for identifying interactions in a larger mass of
target material than is visible to MINERvA and beams that are more
intense and with higher energy neutrinos than NuMI because the beam
becomes narrower and the neutrino interaction rate increases with
neutrino energy.  Of particular interest are the
largest detectors, e.g., IceCube,\cite{Halzen} that uses the Antarctic
icepack to detect events, along with muon storage rings to produce
directed neutrino beams.\cite{huber,Geer}

In summary, we have used the Fermilab NuMI neutrino beam, together with
the \minerva\ detector to provide a demonstration of the possibility
for digital communication using neutrinos.  An overall data rate of
about 0.1 Hz was realized, with an error rate of less than 1\% for
transmission of neutrinos through a few hundred meters of rock.  This
result illustrates the feasibility, but also shows the significant
improvements in neutrino beams and detectors required for practical
applications.

\section*{Acknowledgments}
This work was supported by the Fermi National Accelerator Laboratory,
which is operated by the Fermi Research Alliance, LLC, under contract
No. DE-AC02-07CH11359, including the \minerva\ construction project,
with the United States Department of Energy. Construction support also
was granted by the United States National Science foundation under NSF
Award PHY-0619727 and by the University of Rochester.  Support for
participating scientists was provided by NASA, NSF and DOE (USA) by
CAPES and CNPq (Brazil), by CoNaCyT (Mexico), by CONICYT (Chile), by
CONCYTEC, DGI-PUCP and IDI-UNI (Peru), by Latin American Center
for Physics (CLAF) and by FASI(Russia). Additional support came from Jeffress Memorial
Trust (MK), and Research Corporation (EM).  Finally, the authors are
grateful to the staff of Fermilab for their contribution to this
effort, in particular to Jim Hylen for his tireless support of the
NuMI neutrino beamline.

\end{document}